Tissue Spatial Geometrics Laboratory

BERKELEY LAB
Lawrence Berkeley National Laboratory

# Translating the Architectural Complexity of the Colon or Polyp into a Sinusoid Wave for Classification via the Fast Fourier Transform

David H. Nguyen, PhD
Principal Investigator
Tissue Spatial Geometrics Lab

Affiliate Scientist
Dept. of Cellular & Tissue Imaging
Division of Molecular Biophysics and Integrated Bioimaging
Lawrence Berkeley National Laboratory
DHNguyen@lbl.gov

Tissue Spatial Geometrics Laboratory

BERKELEY LAB
Lawrence Berkeley National Laboratory

## Abstract

**Background**: There is no method to quantify the spatial complexity within colon polyps. This paper describes a spatial transformation that translates the tissue architecture within a polyp, or a normal colon lining, into a complex sinusoid wave composed of discrete points. This sinusoid wave can then undergo the Fast Fourier Transform to obtain a spectrum of frequencies that represents the sinusoid wave. This spectrum can then serve as a signature of the spatial complexity (i.e. an index) within the polyp. **Method**: By overlaying vertical lines that radiate from the bottom-middle (like a fold-out fan) of an image of a polyp stained by hematoxylin & eosin, the image is segmented into sectors. Each vertical line also forms an angle with the horizontal axis of the image, ranging from 0° to 180° rising counter-clockwise. Each vertical line will intersect with various features of the polyp (i.e. border of lumens, border of epithelial lining). Each of these intersections is a point that can be characterized by it's distance from the origin (this distance is also a magnitude of that point). Thus, each intersection between radial line and polyp feature can be mapped by polar coordinates (radius length, angle measure). By summing the distance of all points along the same radial line, each radial line that divides the image becomes one value. Plotting these values (y-variable) against the angle of each radial line from the horizontal axis (x-variable) results in a sinusoid wave consisting of discrete points. This method is referred to as the Linearized Compressed Polar Coordinates (LCPC) Transform. **Discussion**: The LCPC transform, in conjunction with the Fast Fourier Transform, can reduce the complexity of visually hidden histological grades in colon polyps into categories of similar wave frequencies (i.e. each histological grade has a signature consisting of a "handful" of frequencies).

## Graphical Abstract

### Linearized Compressed Polar Coordinate (LCPC) Transform

**Step 1** – Map each border (or feature of interest) using polar coordinates.

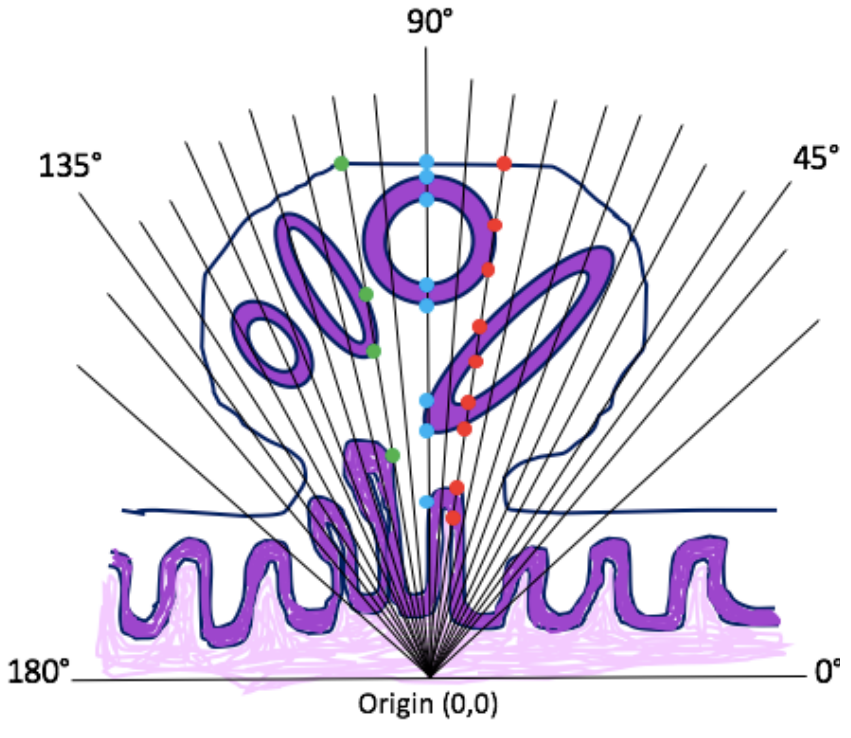

**Step 2** – Compress magnitude of all points on a line into one value for each line.

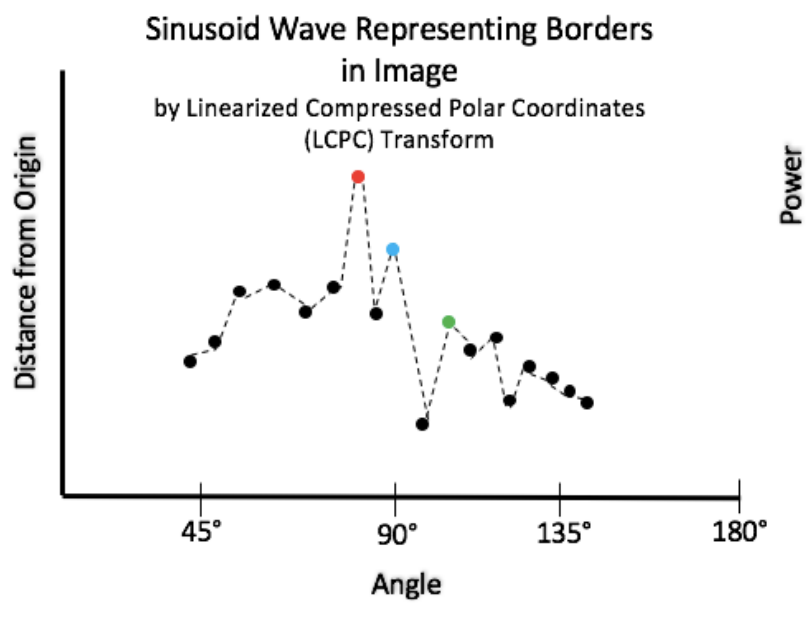

**Step 3** – Apply Fast Fourier Transform

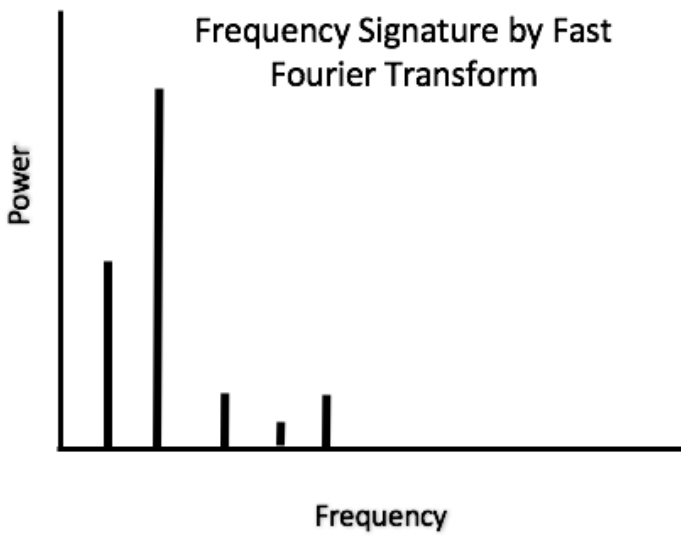

## Applications

- Detect subtle changes in architecture, or cellular components, of colon lining or polyps.
- Classify the complexity of tissue architecture as simple collection of wave frequencies.
- Identify refined polyp grading system to predict clinical outcome or response to therapy of new polyp subtypes.

# Figure 1 – Definition of Method: Linearized Compressed Polar Coordinates

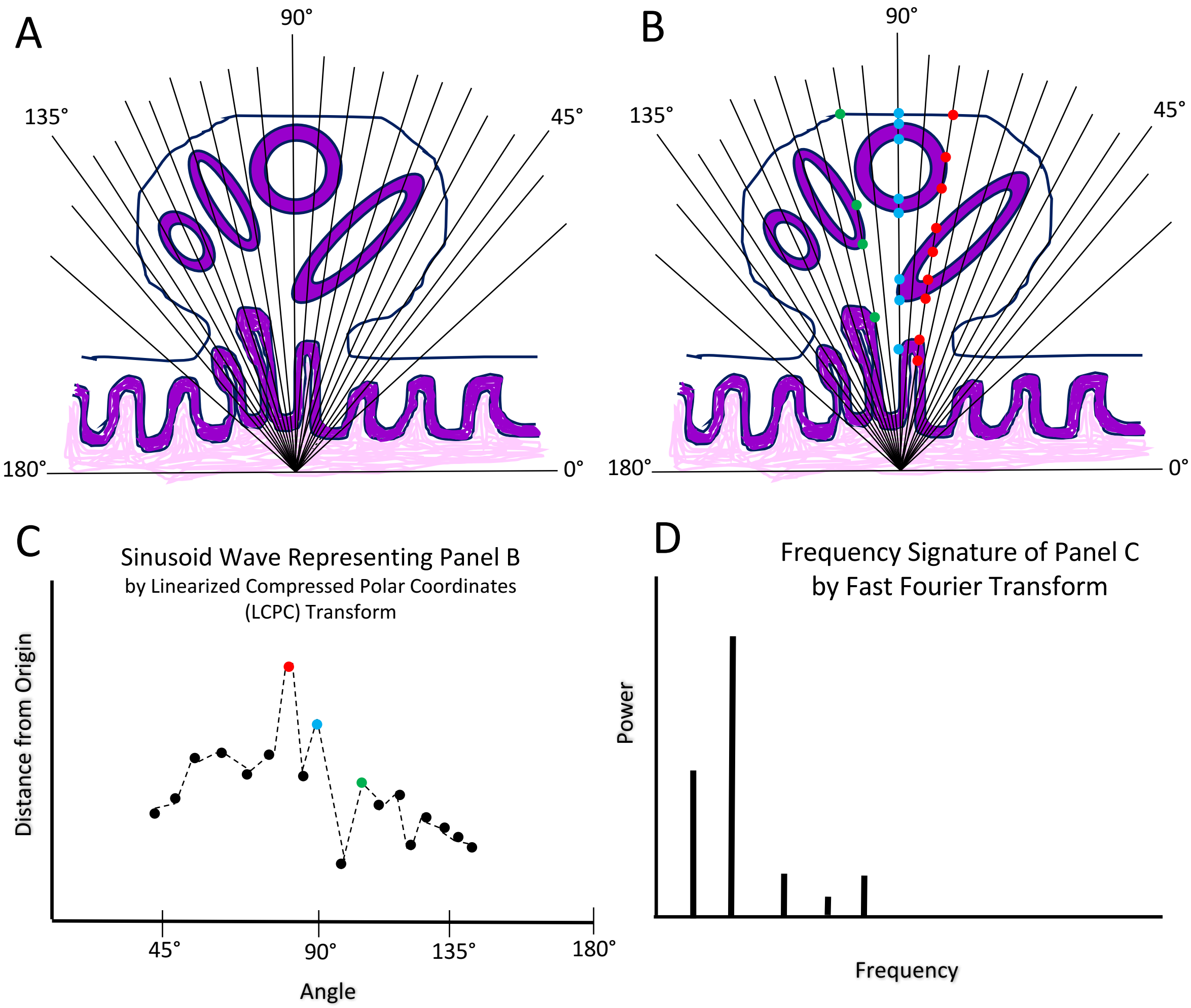

**Figure 1. A Hypothetical Case of Quantifying Tissue Architecture by the LCPC Transform.** (A) Colon polyps protrude above the normal colon lining. (B) A radial grid can segment the polyp and provide points of intersection between each line and the borders of tubular structures within the polyp. (C) By compressing ("adding together") all points of intersection along the same line, each line can be represented by one number. Plotting these numbers versus the angle measure of each corresponding line results in a sinusoid wave consisting of discrete points (dotted line added to accentuate sinusoid shape). (D) The sinusoid wave in C can be represented as a collection of wave frequencies via the Fast Fourier Transform. Thus, the complexity of the tubular structures in A can be represented in the simple manner of a collection of wave frequencies.

## Method

The polar coordinate system consists of (1) an angle that starts from the horizontal x-axis and rises counter-clockwise, and (2) a line that originates from the vertex of the angle. By placing the vertices of all angles at the bottom-middle of a hematoxylin & eosin stained polyp (see Figure 1A), the polyp can be divided into sectors. Each sector is flanked by radial lines whose angles range from 0 to 180 degrees, starting at the horizontal x-axis.

Each radii will intersect features of the polyp, such as individual cells, the inner walls of lumens, and the outer walls of tubular structures (Figure 1B). Each point of intersection is a vector that has a magnitude (the distance from the vertex to the point). By summing the magnitudes of all points on the same line, each line has one numerical value representing the tissue architecture along its path. This summing step is referred to as “compressing.” Plotting the compressed sums of each line versus the angle measure of each line results in a complex sinusoid wave consisting of discrete points (Figure 1C). Since this plot represents angle measures from 0 to 180 degrees along a straight horizontal x-axis, this graph is referred to as a Linearized Compressed Polar Coordinates (LCPC) Transform. The informal name of this spatial transformation is the Nguyen-Nichols Transform, named after Duane Nichols (see Acknowledgements).

As with other sinusoid data, the Fast Fourier Transform (FFT) can be applied to the LCPC plot to obtain the composite frequencies that comprise frequencies compose the complex sinusoid in the LCPC plot (Yoganathan, 1976).

How many segments by which to divide an image?

This number is arbitrary within the following constraints. Too few segments between 0° to 180° will not adequately capture the complexity of tissue architecture, while too many segments unnecessarily creates more work. Sectioning an image by every 1° or 2° may suffice, but this decision should be systematically applied to all samples within and across data sets.

## Applications

The LCPC Transform is versatile in its ability to capture the spatial arrangement of cells and histological features inside a polyp or colon lining. Not only can the LCPC Transform capture the complexity of a tissue’s outer contour, but it can capture the spatial location of individual cells of interest (i.e. immunohistochemically visualized cell types) within the tissue. The LCPC plot reduces the complexity of a tissue architecture into a small collection of wave frequencies.

The original intention of the LCPC Transform was to develop a refined method to classifying colon polyps or inflamed colons into distinct grades that associate with distinct clinical outcome or response to therapy. The author hypothesizes that the LCPC Transform will allow the FFT to identify subsets of frequency profiles that represent non-obvious structural themes in tissues that correlate with different clinical outcome or biological behavior.

Another application of the LCPC Transform can be utilized apart from the FFT. Using the normal colon lining as an example, the LCPC Transform can quantify differences in inflammatory infiltrates between normal colon and inflamed colon. Normal colon and mildly inflamed colon may have a tissue architecture that is too subtly different to be categorized by the human eye, but the LCPC Transform, when used to detect the outer contour along with inflammatory infiltrates can quantitatively capture the subtly. This can be done by subtracting one LCPC plot from another (i.e. the mildly inflamed colon minus the normal colon) (Figure S1).

**Examples**

Colon polyps are graded according to degree of differentiation, vascular and lymphatic involvement, and pedunculated or sessile morphology (Bujanda, 2010). Any feature that can be distinctly identified can have its geometric parameters measured (i.e. length, width, area, distance from something, etc.).

*Pedunculated Adenoma*
Pedunculated adenomas have a stalk that raises the bulb of the polyp away from the normal colon lining (Figure 2 and 3). The protruding bulb can form a mushroom-like shape that is distinct from the mostly flat mucosal lining of the colon. The LCPC Transform is particularly useful for capturing the tissue architecture of pedunculated polyps.

*Normal Colon*
The mucosal surface of the normal colon is largely flat, though it does have finger-tip-like regions that are curved. When applying the LCPC Transform to the flat regions of the normal colon (Figure 4), it is not recommended to use a single point as the origin by which to measure the magnitudes along the length of the radial lines. For flat regions of the normal colon, a straight line running along the base of the lining should be used as the reference by which to measure the distance of each point (the magnitude) from the base. The reasoning behind this recommendation is that, in the normal colon lining, using a single point as the origin of all radial lines will capture a complexity that only occurs in polyps, and thus would be uninformative when comparing normal colon architecture to polyp architecture.

*Sessile Serrated Adenoma*
Sessile Serrated Adenomas are abnormal growth patterns in the colon lining that are not raised above the normal lining like a mushroom. For these types of polyps, it is recommended that a horizontal line along the base of the polyp lining be used as a reference to measure the distance of each point (the magnitude) from the base (Figure 5).

Tissue Spatial Geometrics Laboratory

BERKELEY LAB
Lawrence Berkeley National Laboratory

## Figure 2 – Tubular Adenoma, Outer Contour

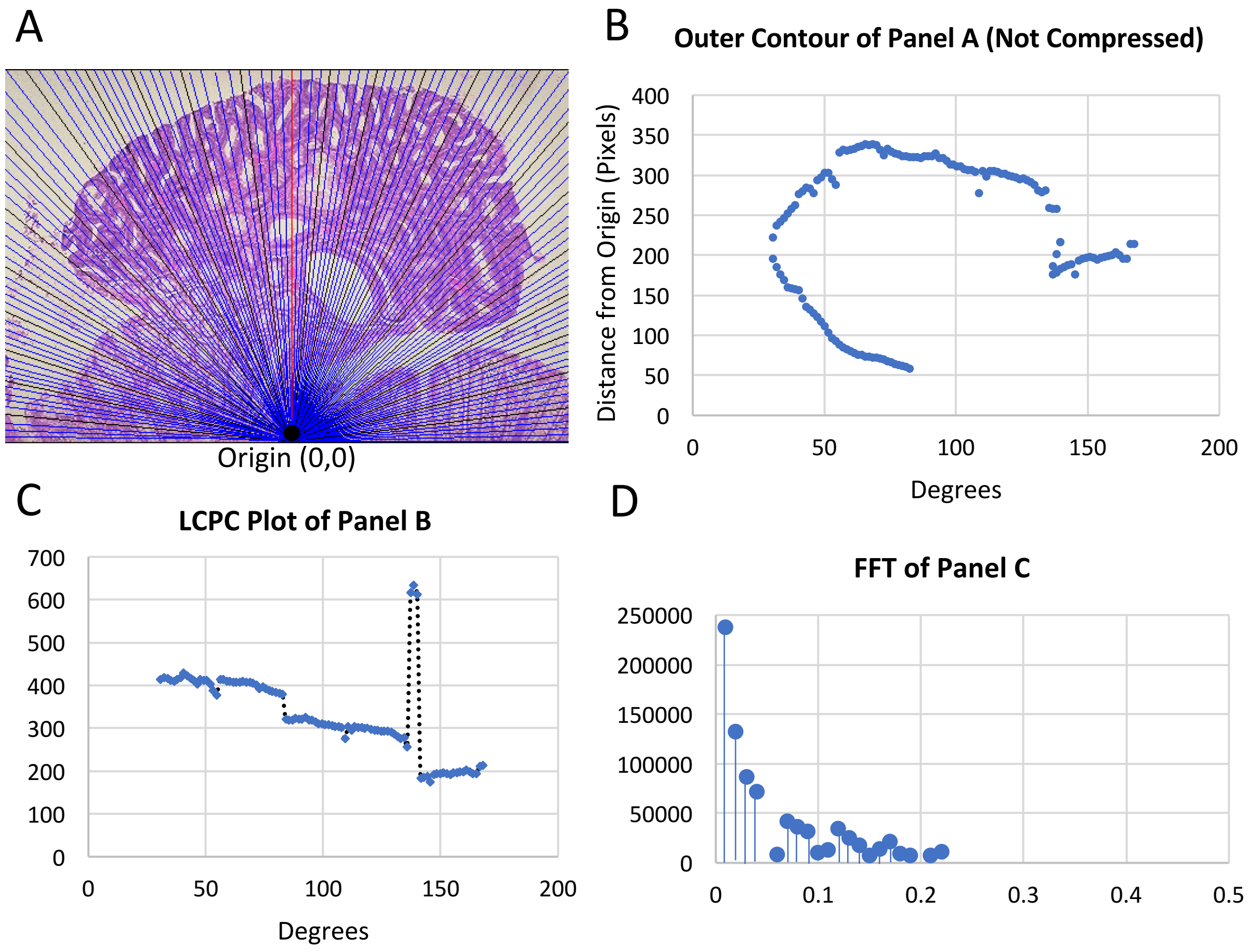

**Figure 2. The LCPC Transform Can Represent the Complexity of a Polyp's Outer Contour.** The LCPC Transform can quantify the complexity of a polyps outer surface, which may be of value for diagnostic purposes or for detecting differential tissue behavior in biomedical research. (A) The radial segments intersect with the outer surface of the polyp. Some lines intersect the outer contour more than once. (B) A plot of the outer contour based on points of intersection with the radial lines. (C) By summing the magnitude of all points ("compressing") along the same line, the outer contour of the polyp can be represented by a sinusoid wave consisting of discrete points (dotted line added to accentuate sinusoid shape). (D) The sinusoid can be transformed into a collection of wave frequencies via the Fast Fourier Transform. This collection of frequencies represents the complexity of the outer contour.

# Figure 3 – Tubular Adenoma, All Outer and Luminal Borders

A

B

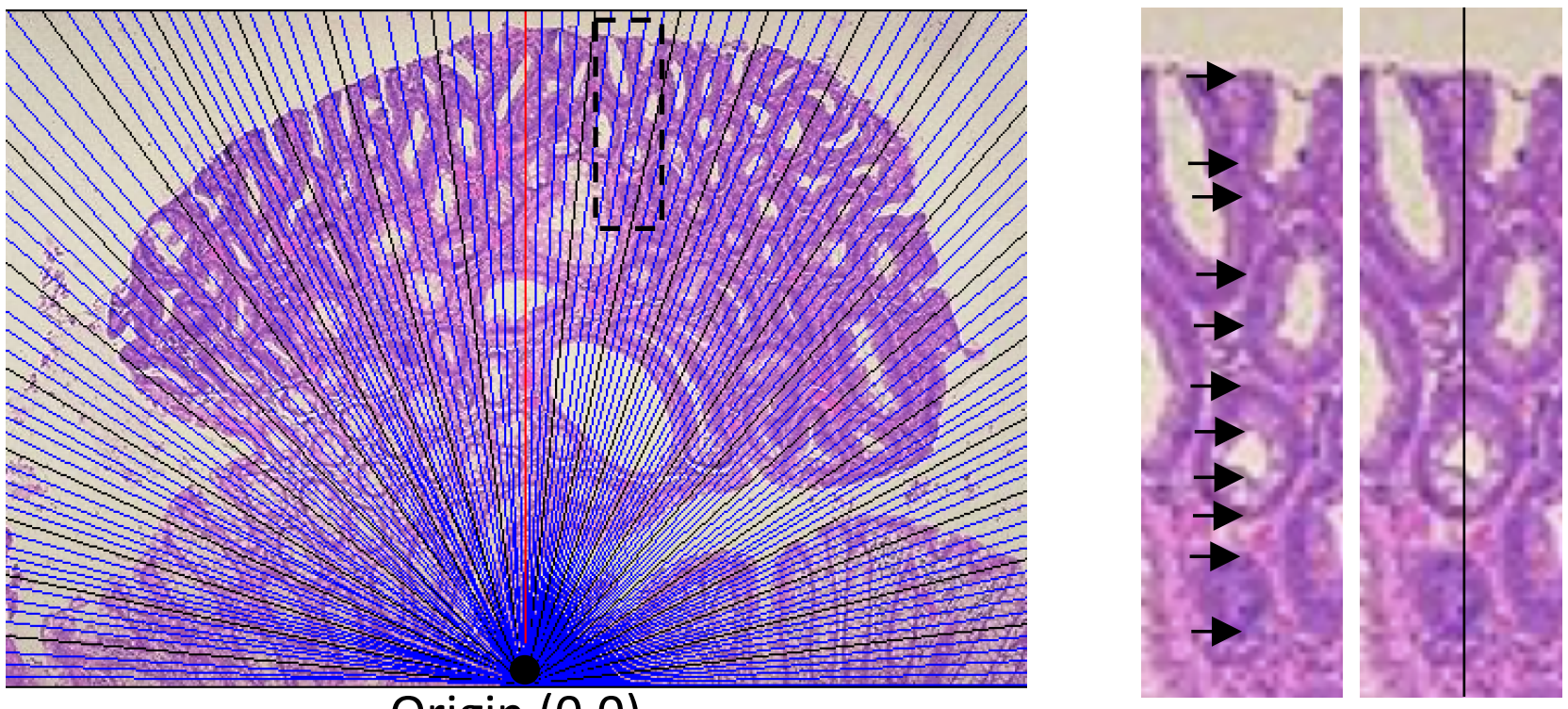

C

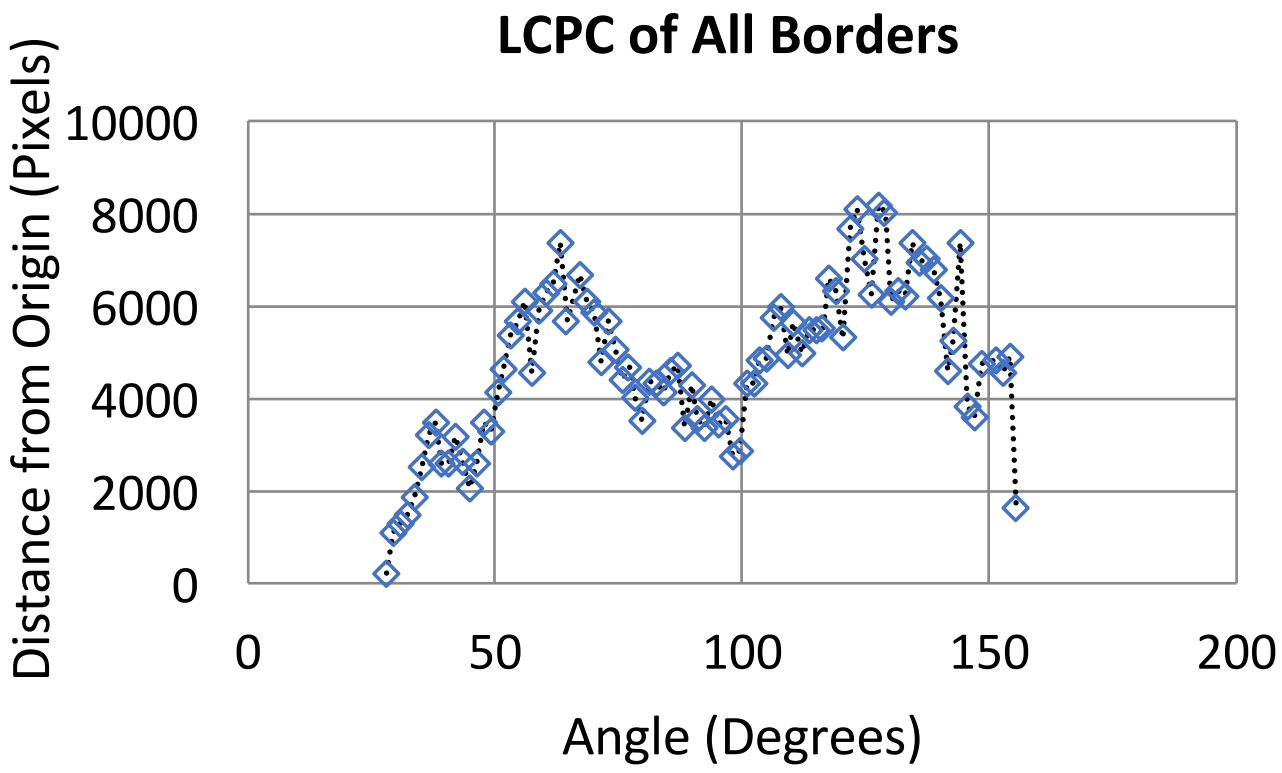

D

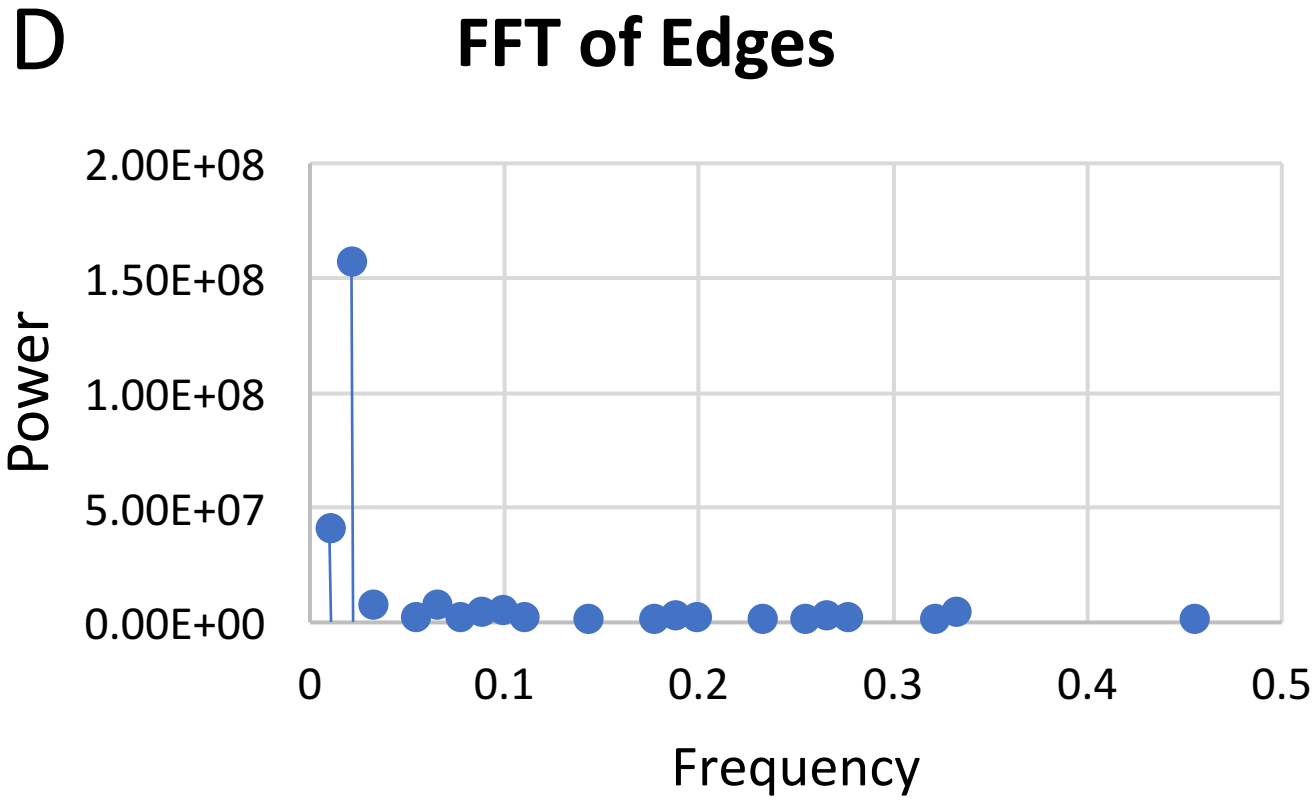

**Figure 3. LCPC Transform of Tubular Structures in a Tubular Adenoma.**
(A) Tubular Adenomas are pedunculated, meaning a stalk raises them up like a mushroom. Placing the origin at the bottom-middle of the image defines a vertex for all angles that segment the image from 0° to 180°. (B) Inset of A. Example of tubular borders that are counted. (C) The LCPC Transform of the image in A. (D) The Fast Fourier Transform of the sinusoid wave in C.

Figure 4 – Normal Colon

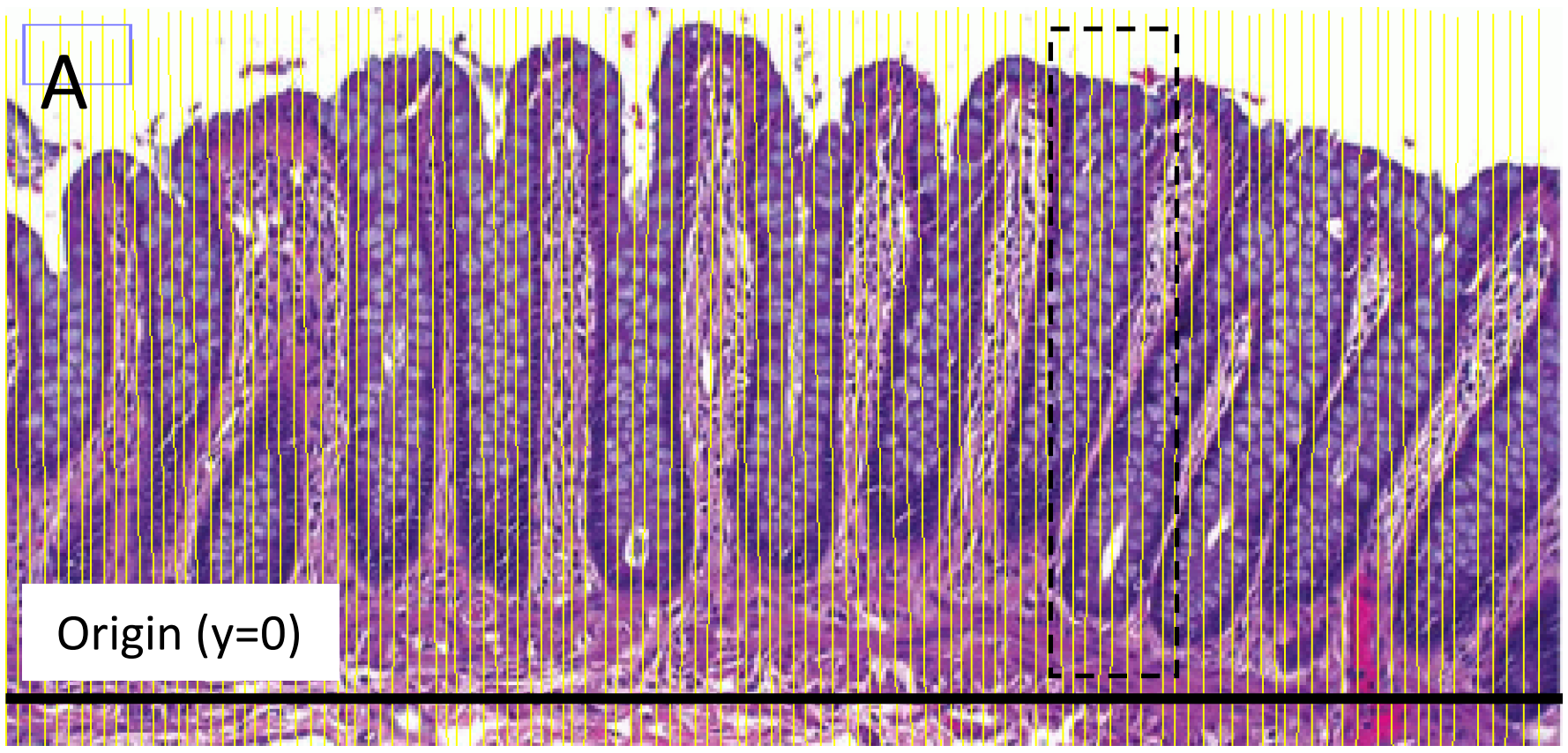

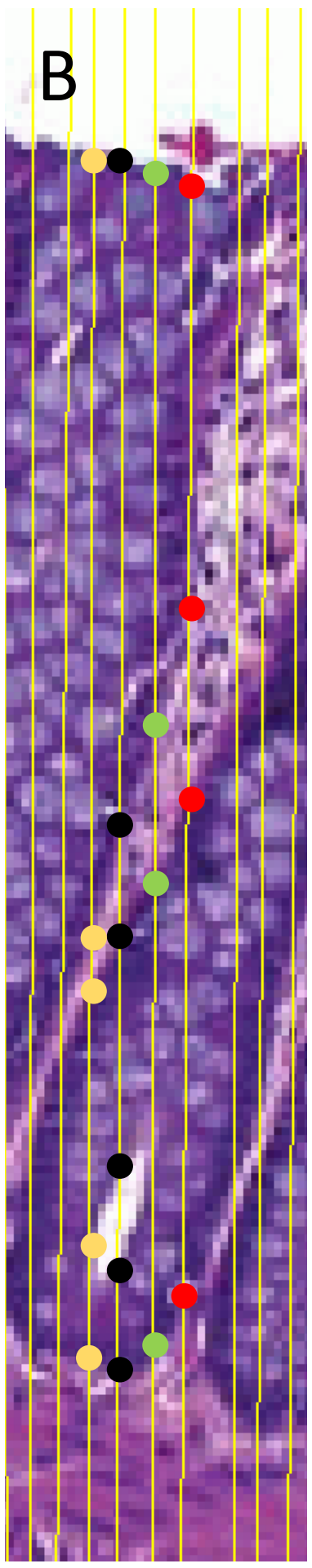

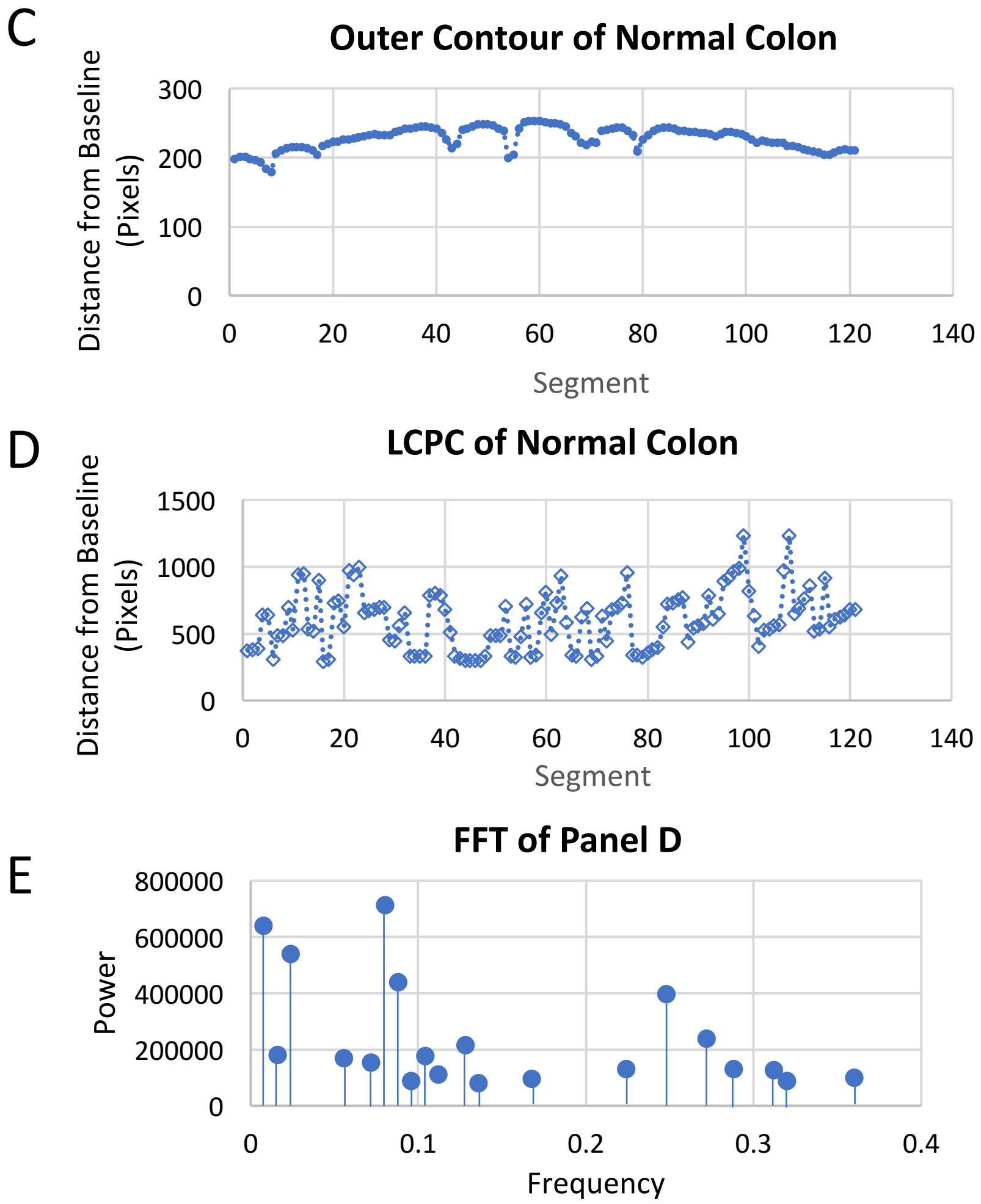

**Figure 4. LCPC Transform of the Normal Colon.** (A) A line along the base of the lining is the reference by which to measure the magnitude to points. (B) Inset of A. Example tubular edges counted. (C) Outer contour of A. (D) LCPC Transform of A (dotted line added to accentuate sinusoid shape). (E) The Fast Fourier Transform of D.

# Figure 5 – Sessile Serrated Adenoma

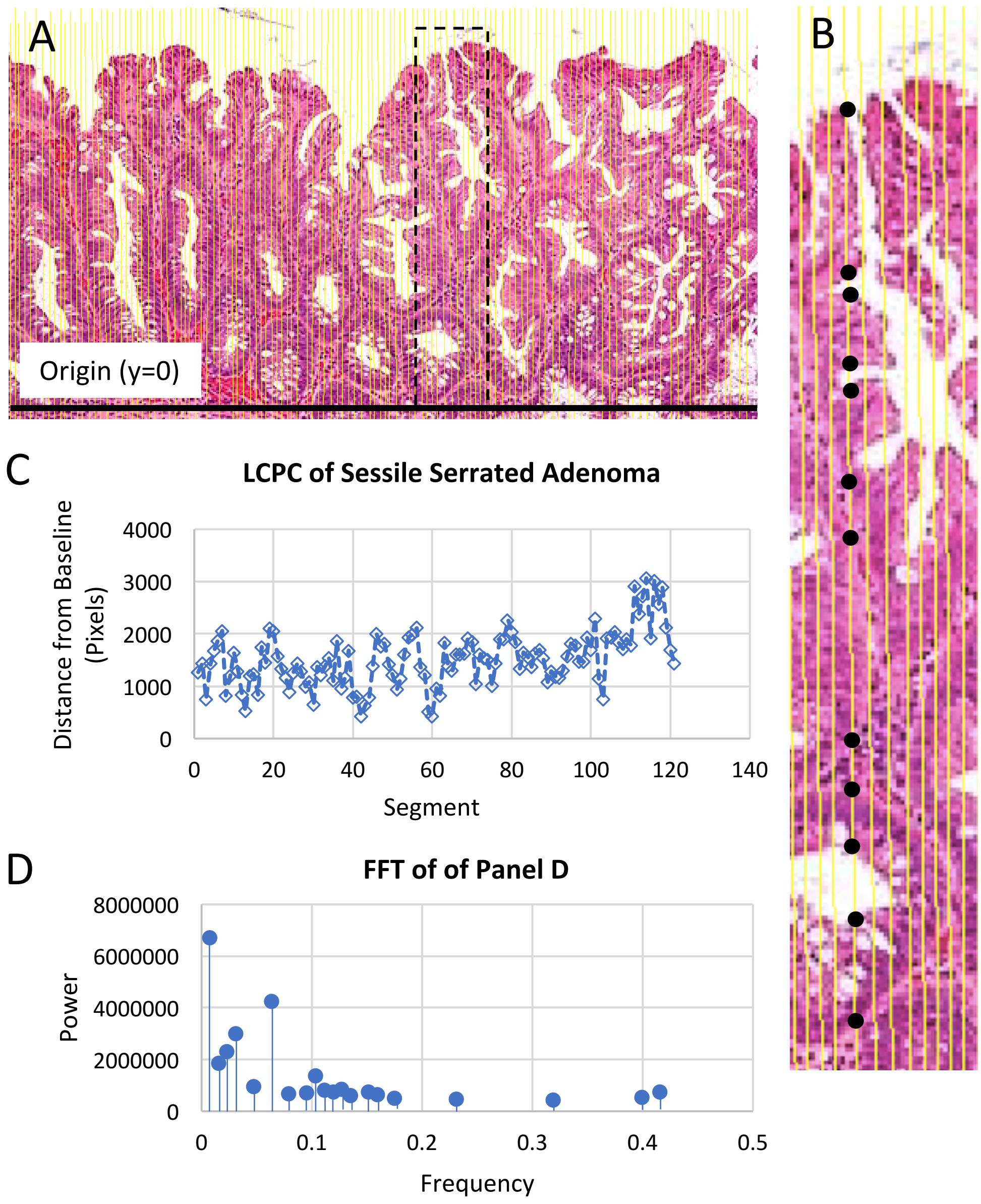

**Figure 5. The LCPC Transform on Sessile Serrated Polyps**

(A) Sessile serrated polyps are best characterized by the LCPC Transform via vertical segments that extend from the same baseline. (B) Inset in A. Example of borders that were counted. (C) LCPC plot of the image in A. (D) Fast Fourier Transform of Panel C.

# Figure S1

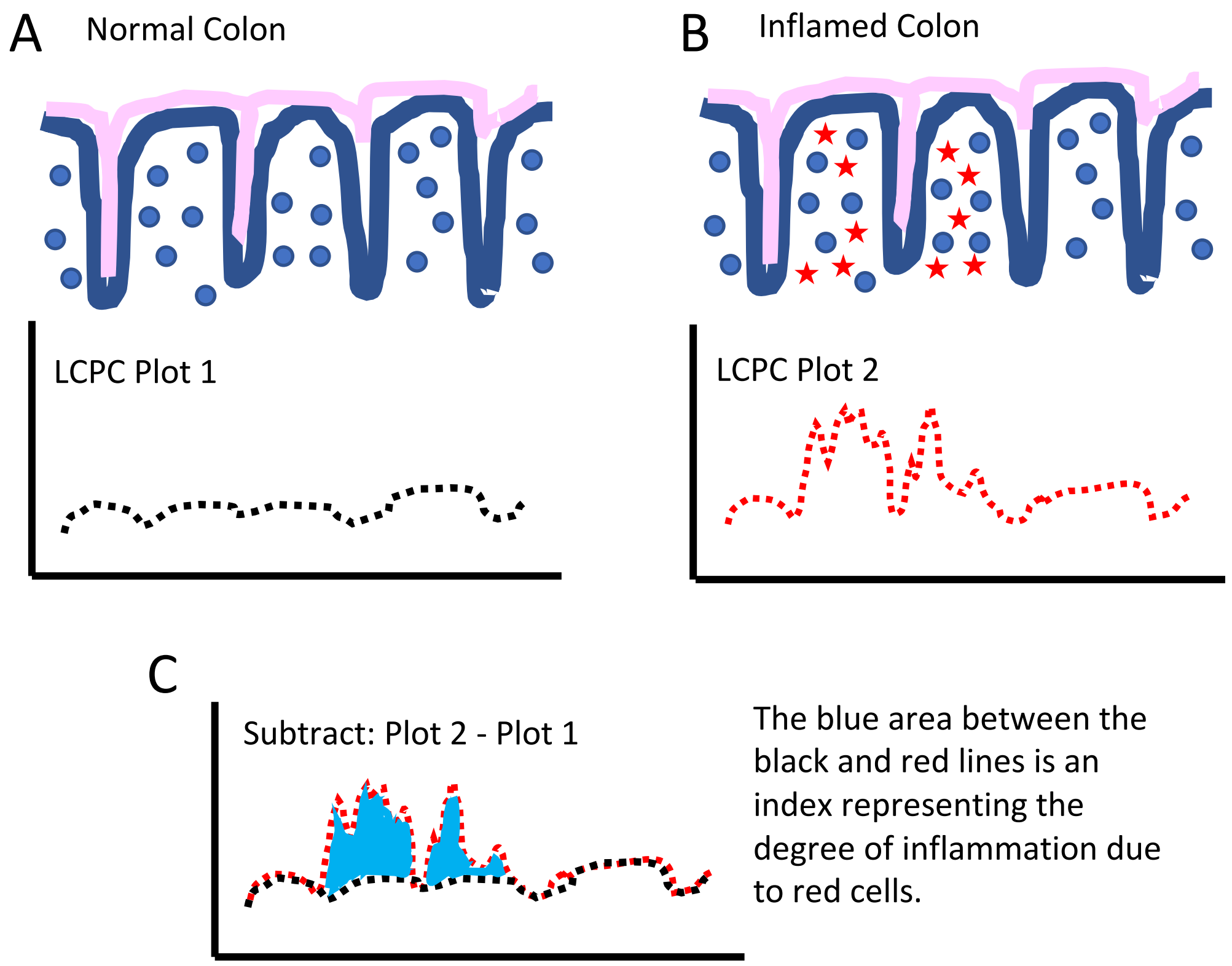

**Figure S1. The LCPC Transform Can Quantify the Degrees of Inflammatory Infiltration** Inflamed colon lining exhibits a higher number of inflammatory cells compared to normal colon. The width of the folds may also be larger in the inflamed colon, which will change the shape of the LCPC plot compared to normal colon. (A) Normal colon with usual number of inflammatory cells; along with its hypothetical LCPC plot. (B) Inflamed colon exhibiting additional inflammatory cell types (red stars); along with its hypothetical LCPC plot. (C) By subtracting LCPC plots between different conditions, the features of interest that were measured in obtaining the LCPC plot can be quantified. In this case, the complexity added by the red stars can be represented as the numerical value equated with the area between the two curves.

**Acknowledgements**

The author would like to dedicate this algorithm in memory of Mr. Duane Nichols, M.S., who taught biology at Alhambra High School (Alhambra, CA). Mr. Nichols passed away in 2017 because of colon cancer, but left behind 30+ years of students inspired to do STEM research.